\documentclass[11pt]{article}

\usepackage[T1]{fontenc}
\usepackage[utf8]{inputenc}
\usepackage[a4paper,margin=1in]{geometry}
\usepackage{amsmath,amssymb}
\usepackage{array}
\usepackage{authblk}
\usepackage{booktabs}
\usepackage{graphicx}
\usepackage{float}
\usepackage{placeins}
\usepackage[protrusion=true,expansion=false]{microtype}
\usepackage{xcolor}
\usepackage[colorlinks=true,linkcolor=blue!45!black,citecolor=blue!45!black,urlcolor=blue!45!black]{hyperref}

\definecolor{geoblue}{HTML}{3867A8}
\definecolor{geoteal}{HTML}{2A9D8F}
\definecolor{geocoral}{HTML}{C85A54}
\definecolor{geogray}{HTML}{F2F5F8}
\definecolor{geotext}{HTML}{243040}

\newcommand{\method}{GeoContra}
\newcommand{\llmonly}{LLM-only}
\newcommand{\ours}{\method}
\newcommand{\ind}{\mathbb{I}}
\newcommand{\pp}{\,pp}
\newcommand{\rate}[1]{#1\%}
\newcolumntype{P}[1]{>{\raggedright\arraybackslash}p{#1}}
\newcolumntype{T}[1]{>{\vspace{0pt}\raggedright\arraybackslash}p{#1}}
\newcounter{algorithmblock}
\newenvironment{algorithmblock}[1]{%
  \refstepcounter{algorithmblock}%
  \par\medskip
  \noindent\begin{minipage}{\linewidth}
  \hrule\vspace{5pt}
  \noindent\textbf{Algorithm \thealgorithmblock. #1}\par\vspace{4pt}
  \begin{tabular}{@{}T{0.12\linewidth}T{0.82\linewidth}@{}}
}{%
  \end{tabular}
  \vspace{4pt}\hrule
  \end{minipage}
  \par\medskip
}

\title{\textbf{\method: From Fluent GIS Code to Verifiable Spatial Analysis with Geography-Grounded Repair}\thanks{This work was supported by the Guangdong Basic and Applied Basic Research Foundation (Grant No. 2025A1515010111, 2025A1515010513), Guangdong Natural Science Foundation (2024A1515012080), and the Third Batch of Special Entrusted Projects of Guangdong Philosophy and Social Sciences Innovation Project (GD25WTCXGC14).}}
\author[1]{Yinhao Xiao\thanks{\texttt{20191081@gdufe.edu.cn}}}
\author[2,3]{Rongbo Xiao\thanks{Corresponding author. \texttt{ecoxiaorb@163.com}}}
\author[2,3,4]{Yihan Zhang\thanks{Corresponding author. \texttt{zyh4184@gdufe.edu.cn}}}
\affil[1]{School of Big Data and Artificial Intelligence, Guangdong University of Finance and Economics}
\affil[2]{School of Geography and Environment Economics, Guangdong University of Finance \& Economics}
\affil[3]{Guangdong Engineering Research Center of Low-Altitude Remote Sensing Intelligent Monitoring}
\affil[4]{Research Center for Eco-Product Accounting and Application, Guangdong University of Finance \& Economics}
\date{}

\begin{document}
\maketitle

\begin{abstract}
Reliable spatial analysis is a challenging task in GIScience. An analysis must preserve coordinate-reference semantics, spatial predicates, topology, measurement units, spatial support, and geographic plausibility. The current gap is that LLM-based GIS systems increasingly generate fluent scripts and tool workflows, but they rarely make these geographic rules explicit or verify them at scale on real local GIS data. We present \method, a geography-grounded verification and repair framework for LLM-driven Python GIS workflows. \method{} represents each task as an executable geospatial contract over a natural-language spatial question, dataset schemas, CRS metadata, expected outputs, spatial relation semantics, topology and metric constraints, required GIS operations, and forbidden shortcuts. Generated programs are checked by static geographic-rule inspection, runtime artifact validation, and task-level semantic verifiers; violations are returned to the model in a bounded repair loop. We evaluate on \textbf{7,079} real geospatial tasks spanning \textbf{15} Greater Boston areas and \textbf{9} GIS task families, plus an \textbf{11-model} open-source suite with \textbf{600} runs per model. To our knowledge, this is the first large-scale benchmark built specifically around executable, contract-verifiable LLM GIS workflows. On the full closed-model benchmark, \method{} improves spatial correctness from \rate{47.6} to \rate{77.5} for DeepSeek-V4 and from \rate{57.7} to \rate{81.5} for Kimi-K2.5. Across the 11 open models, average spatial correctness rises by \textbf{26.6} percentage points. The significance is methodological: \method{} turns LLM GIS automation from fluent code production into verifiable spatial analysis by catching negative travel times, CRS and field-schema violations, missing spatial predicates, and brittle output casts that otherwise produce executable but geographically invalid results.
\end{abstract}


\section{Introduction}

The central challenge of automated GIS is not merely whether a system can write code, but whether it can preserve the geographic meaning of an analysis. Traditional GIS scripting is already error-prone because analysts must translate a spatial question into a chain of operations whose validity depends on coordinate reference systems, measurement units, spatial support, geometry validity, topology, spatial predicate direction, raster resolution, graph connectivity, and output interpretation. A script may be syntactically correct while buffering in degrees, aggregating over the wrong areal unit, dropping zero-count source objects, treating a Euclidean distance as network accessibility, or returning a negative travel time. Such failures undermine reproducibility and spatial decision support because they change the geographic question being answered.

Large language models intensify this long-standing GIS problem. Recent study on GeoFM argues that foundation models may reshape spatial data science and GeoAI, but also raises challenges around geo-alignment, multimodality, and spatially explicit evaluation \cite{janowicz2025geofm}. Nature Machine Intelligence has similarly emphasized that geospatial foundation models should be developed responsibly rather than only scaled for performance \cite{natmi2025responsible}, and recent Earth-observation foundation models show the speed at which geospatial AI is becoming more general and multimodal \cite{wu2025skysensepp}. In parallel, recent GIS and geoscience LLM work has adapted models to domain knowledge, GIS tools, topological reasoning, and geospatial code generation \cite{deng2024k2,zhang2024bbgeogpt,wei2024geotool,ji2025georeasoning,hou2025geocogent,lin2026geoagent}; other benchmarks evaluate geospatial code generation and Google Earth Engine programs \cite{gramacki2024evaluation,hou2025canllm,wu2025autogeevalpp}. These efforts demonstrate the promise of LLM-based GIS automation, but they also reveal a gap: most systems remain centered on tool invocation, code fluency, foundation-model capability, or platform-specific execution, while the GIScience question of \emph{geographic rule preservation} is only partially checked. In particular, existing evaluations rarely combine real local vector, raster, network, and topology data with executable contracts that can test CRS correctness, spatial-relation semantics, topology consistency, metric plausibility, and row-level reproducibility at scale. Closing this gap matters because an LLM-GIS workflow that is executable but geographically wrong can silently distort accessibility, exposure, environmental justice, and planning conclusions.

This paper takes a geography-driven position: LLM-driven GIS should be evaluated and repaired according to explicit spatial contracts, not only by whether generated code runs. We introduce \method, a framework that wraps an LLM with machine-checkable geospatial contracts and a verifier-guided repair loop. The model still produces Python GIS workflows, but \method{} specifies the spatial question, data representations, expected output, required geographic operations, forbidden shortcuts, and geographic validity rules before generation. After generation, the system checks whether the workflow preserves geodetic, topological, relational, metric, and tabular semantics, and it returns concrete geographic violations to the model for repair.

The empirical setting is deliberately broad. The benchmark contains 7,079 real tasks generated from local geospatial layers for 15 Greater Boston areas, covering buffer counting, spatial joins, nearest-neighbor distances, network accessibility, overlay areas, raster sampling, zonal statistics, topology-quality tasks, and multi-step workflows. We run paired \llmonly{} and \ours{} evaluations for DeepSeek-V4 and Kimi-K2.5 on the full benchmark, yielding 14,158 rows per model family. We further evaluate 11 open models, each on 300 selected tasks under two settings, yielding 600 runs per model.

The benchmark is convincing because it is not a synthetic toy suite of isolated prompts. It is built from real local geospatial assets, mixes vector, raster, network, and topology operations, spans municipalities with heterogeneous feature distributions, and encodes each task as an executable contract with schema, CRS, method, topology, metric, and output requirements. In practice, industry GIS data are often private, fragmented, and not released with referenceable task contracts, which leaves LLM GIS evaluation without a shared, reproducible stress test. To the best of our knowledge, \textsc{GeoContra-Real} is the first benchmark at this scale to make executable geospatial contracts central to evaluating and repairing LLM-driven GIS workflows.

The contributions are:
\begin{itemize}
  \item A geography-grounded formulation of LLM-driven spatial analysis, where contracts encode CRS validity, spatial-relation semantics, topology, metric units, spatial support, and output reproducibility.
  \item A scalable real-world GIS benchmark, \textsc{GeoContra-Real}, with 7,079 tasks across 15 urban areas and 9 task families; to our knowledge, it is the first large-scale benchmark designed around executable, contract-verifiable LLM GIS workflows.
  \item A multilayer verification and bounded repair framework that translates GIScience rules into static, runtime, and semantic checks, then uses geographic violations rather than generic self-reflection to repair workflows.
  \item An extensive evaluation showing consistent improvements on closed and open models, accompanied by case studies that explain how contract feedback restores geographic fidelity.
\end{itemize}

\section{Related Work}
\label{sec:related-work}

Recent GIScience scholarship increasingly treats foundation models and LLM agents as possible infrastructure for next-generation GeoAI. Janowicz \emph{et al.} frames GeoFM as a shift that could reshape spatial data science research, education, and practice, while emphasizing geo-alignment, multimodal geospatial representation, and evaluation as open challenges \cite{janowicz2025geofm}. K2 and BB-GeoGPT show parallel attempts to adapt language models to geoscience and GIS knowledge through domain corpora, instruction tuning, and domain-specific benchmarks \cite{deng2024k2,zhang2024bbgeogpt}. Ji \emph{et al.} show that LLMs can partially infer topological spatial relations from vector geometries and language, but that spatial relation reasoning remains limited and representation-dependent \cite{ji2025georeasoning}. In Nature Machine Intelligence, both the editorial on responsible geospatial foundation models and SkySense++ illustrate a broader movement toward large-scale geospatial AI, while also highlighting sustainability, modality, and generalization challenges \cite{natmi2025responsible,wu2025skysensepp}. These works motivate our view that the relevant target is not generic automation but geographically meaningful and accountable analysis.

A growing number of GeoAI studies LLMs for geospatial tool use, workflow construction, and code generation. GeoTool-GPT fine-tunes LLMs to understand GIS tools and construct tool-use solutions \cite{wei2024geotool}. GeoCogent integrates planning, tool-augmented reasoning, memory, and code generation for geospatial programming \cite{hou2025geocogent}. GeoAgent develops a hierarchical multi-agent architecture for autonomous spatial analysis with planning, execution, and review layers \cite{lin2026geoagent}. ChatGeoAI and GeoGPT translate natural-language requests into executable GIS or PyQGIS workflows for non-expert and professional users \cite{mansourian2024chatgeoai,zhang2024geogpt}; MapGPT extends the same agentic idea to cartographic design and map-element control \cite{zhang2024mapgpt}; and a flood knowledge-constrained LLM demonstrates how GIS interaction can be grounded in domain knowledge for risk communication \cite{zhu2024flood}. Outside these agent systems, GeoCode-GPT builds geospatial code corpora and a domain-specific LLM for geospatial code generation \cite{hou2025geocodegpt}. These works improve the ability of models to select tools, write code, and follow geospatial programming conventions. However, tool-use correctness and code executability do not by themselves guarantee that a spatial analysis preserves CRS semantics, topological validity, spatial predicate meaning, metric realism, or the intended spatial support. \method{} is therefore complementary: instead of training a new model or adding another agent planner, it wraps any model with geography-grounded contracts and verifies whether the generated workflow answers the intended spatial question.

Benchmarking and evaluation are also moving quickly. Gramacki \emph{et al.} evaluate code LLMs on manually designed geospatial Python tasks with executable tests \cite{gramacki2024evaluation}, and Zhang and Gao demonstrate ChatGPT-4 for geospatial workflow automation in SIGSPATIAL experiments \cite{zhang2024chatgpt4}. The GeoCode-Eval study asks whether LLMs can generate geospatial code and evaluates models across cognition, interpretation, and creation dimensions \cite{hou2025canllm}. AutoGEEval++ provides a multi-level, multi-modality automated benchmark for Google Earth Engine code generation, with thousands of tests across GEE data types \cite{wu2025autogeevalpp}. These benchmarks are valuable, but they are often platform-specific, emphasize code correctness or operator knowledge, or do not jointly stress local vector, raster, network, topology, and multi-step GIS workflows. \textsc{GeoContra-Real} differs by using thousands of executable real local GIS tasks and embedding verifiable contracts that check geographic invariants rather than only syntactic or API-level success.

Finally, \method{} connects to reproducible and trustworthy spatial analysis. Whereas many LLM-GIS systems treat verification as debugging after generation, our work treats verification as a formal part of the spatial-analysis design. The contract-checking mechanism enforces spatial support, predicate semantics, topology, CRS and unit consistency, and metric plausibility during generation and repair, thereby moving LLM-driven GIS from code assistance toward verifiable spatial analysis.

\section{Problem Formulation}
\label{sec:formulation}

Let a geospatial task be a tuple
\begin{equation}
  \tau = (q, \mathcal{D}, \mathcal{Y}, \mathcal{C}, \mathcal{M}^{+}, \mathcal{M}^{-}),
\end{equation}
where $q$ is the natural-language request, $\mathcal{D}$ is the set of input datasets with schemas and coordinate reference systems, $\mathcal{Y}$ specifies the expected output path, columns, row-count constraints, and metric ranges, $\mathcal{C}$ is a set of spatial constraints, $\mathcal{M}^{+}$ is a set of expected methods, and $\mathcal{M}^{-}$ is a set of forbidden shortcuts. A model $G_{\theta}$ maps a prompt representation $\phi(\tau)$ to a program $p$:
\begin{equation}
  p_0 = G_{\theta}\!\left(\phi(\tau)\right).
\end{equation}

The tuple is intended to be geographic rather than merely computational. The dataset set $\mathcal{D}$ represents observations of geographic phenomena at particular spatial supports and coordinate frames; $\mathcal{C}$ encodes geographic validity rules such as projected CRS requirements for metric operations, topology validity, spatial predicate semantics, ratio bounds, non-negative distances and travel times, and preservation of source spatial units; $\mathcal{M}^{+}$ and $\mathcal{M}^{-}$ distinguish legitimate GIS operations from shortcuts that answer a different spatial question. Thus, a task contract specifies not only what code should be produced, but also which geographic interpretation must be preserved.

We define three verifier families. The static verifier $V_s$ inspects the program before trusting its result, including CRS-unit checks, spatial-predicate direction, schema-field references, topology-sensitive operation patterns, and forbidden method patterns. The runtime verifier $V_r$ executes the program and checks whether the produced artifact satisfies geographic output contracts: expected columns, row-count preservation, duplicate IDs, non-negative metrics, bounded ratios, and readable spatial or tabular files. The task verifier $V_t$ checks lightweight semantic obligations such as requiring a buffer operation in buffer-count tasks, overlay logic in area-ratio tasks, raster sampling in raster tasks, or shortest-path logic in network tasks. The full violation set is
\begin{equation}
  V(p,\tau) = V_s(p,\tau) \cup V_r(p,\tau) \cup V_t(p,\tau).
\end{equation}

A program is spatially correct if it executes and has no remaining geographic-contract violations:
\begin{equation}
  S(p,\tau) =
  \ind\left[\operatorname{exec}(p,\tau)=1 \wedge |V(p,\tau)|=0\right].
\end{equation}
For a model family $m$ and setting $z$, the reported spatial correctness is
\begin{equation}
  R_{m,z} = \frac{1}{|\mathcal{T}|}\sum_{\tau \in \mathcal{T}} S(p_{m,z,\tau},\tau),
  \qquad
  \Delta_m = R_{m,\ours} - R_{m,\llmonly}.
\end{equation}
The repair loop seeks a program with an empty violation set under a bounded interaction budget:
\begin{equation}
  p_{r+1}=G_{\theta}\!\left(\psi(\tau,p_r,V(p_r,\tau))\right),
  \qquad r=0,\ldots,R_{\max}-1.
\end{equation}

\section{\method}
\label{sec:method}

\begin{figure}[t]
  \centering
  \includegraphics[width=\linewidth]{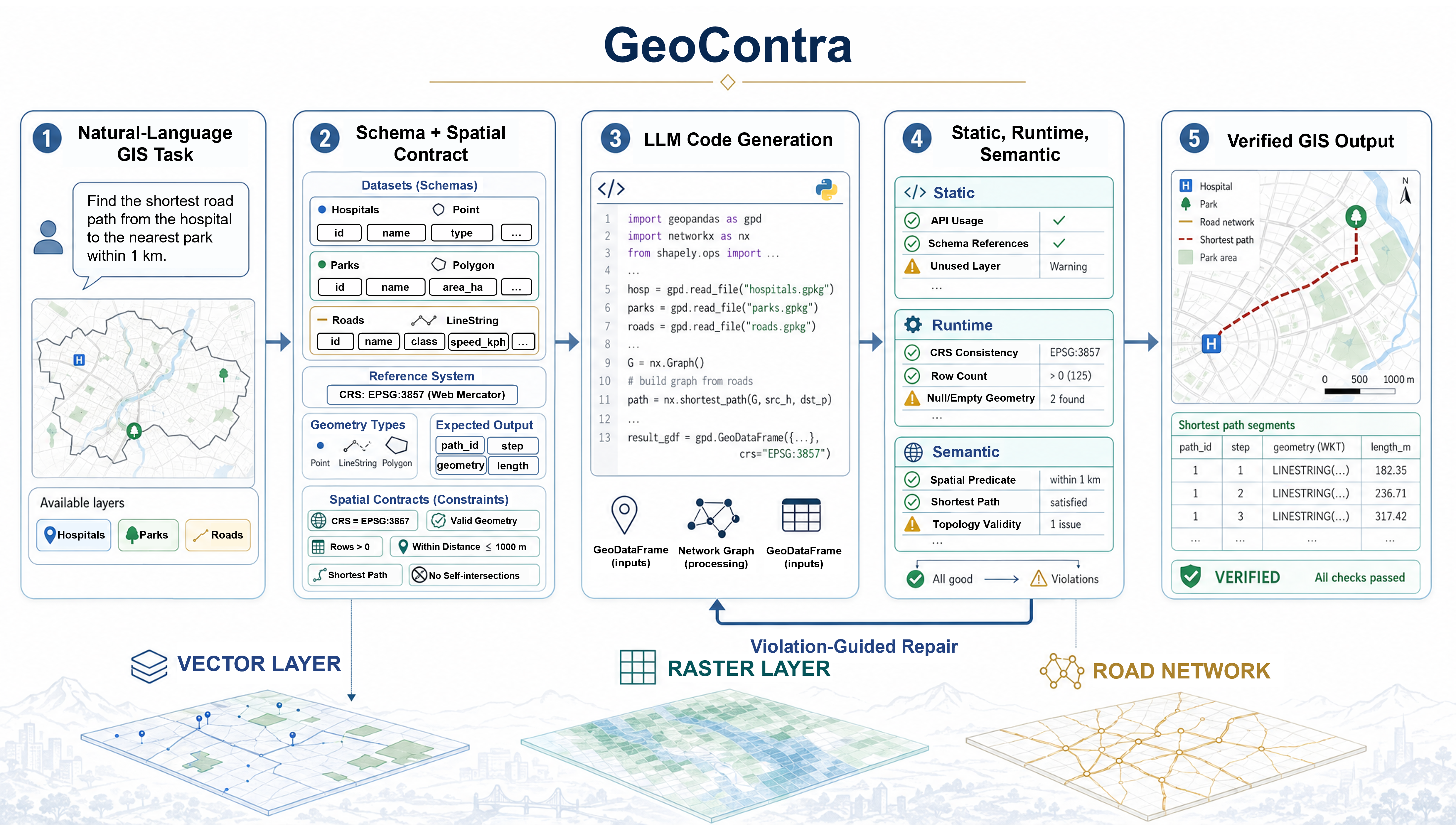}
  \caption{Overview of \method{}. The pipeline begins with a natural-language GIS task and local geospatial data, constructs a geography-grounded task contract containing dataset schemas, CRS information, expected outputs, topology and metric constraints, spatial-relation semantics, expected GIS operations, and forbidden shortcuts, then asks an LLM to generate Python GIS code. The generated program is checked by static, runtime, and semantic verifiers; violation feedback is converted into a repair prompt and returned to the LLM until the program is executable and geographically violation-free or the repair budget is exhausted.}
  \label{fig:overview}
\end{figure}

\method{} has four components: a benchmark/task contract generator, a constraint-aware prompt, a multilayer verifier, and a bounded repair loop. The design goal is to translate GIScience rules into executable constraints that can be checked automatically. Figure~\ref{fig:overview} gives the high-level view: a natural-language GIS request is converted into a schema- and constraint-rich task contract, the LLM produces executable Python GIS code, and the generated program is accepted only after verification or sent back through violation-guided repair. The figure emphasizes that \method{} is not a prompt template alone; it is a closed-loop system in which task contracts, generated code, verifier feedback, and final geospatial artifacts are all explicit objects.

\subsection{Task Contracts}

Each task is stored as a JSON contract. Besides the user query, the contract contains dataset paths, geometry types, CRS metadata, field names, row counts, expected output columns, and GIS-specific constraints. These fields encode the spatial support of the analysis, the coordinate frame in which measurements are meaningful, the source and target roles of layers, and the invariants that the output must preserve. For example, a buffer-count task carries constraints such as \texttt{crs\_projected\_required}, \texttt{one\_output\_row\_per\_source}, and \texttt{nonnegative\_counts}; a network accessibility task includes \texttt{network\_distance\_required}, \texttt{travel\_time\_unit\_minutes}, expected methods such as \texttt{load\_graphml}, \texttt{nearest\_nodes}, and \texttt{shortest\_path}, and forbidden direct-geometry distance shortcuts. In this sense, the contract records the geographic analysis design that the LLM must respect.

\begin{algorithmblock}{Constructing \textsc{GeoContra-Real}}
\label{alg:benchmark}
\textbf{Input} & Area list $\mathcal{A}$, local OSM-derived layers, generation configuration.\\
1 & For each area $a \in \mathcal{A}$, build vector layers, raster layers, walk graph, topology perturbations, and analysis units.\\
2 & Normalize each layer into a schema record containing path, geometry type, CRS, fields, and row count.\\
3 & Enumerate task templates for buffer counts, spatial joins, overlays, nearest neighbors, network accessibility, rasters, topology, and multi-step workflows.\\
4 & Instantiate each template with source/target layers, distance thresholds, $k$ values, and expected output columns.\\
5 & Attach constraints $\mathcal{C}$, expected methods $\mathcal{M}^{+}$, forbidden methods $\mathcal{M}^{-}$, and output contracts $\mathcal{Y}$.\\
\textbf{Output} & A task set $\mathcal{T}$ of JSON contracts and a tabular task index.\\
\end{algorithmblock}

Algorithm~\ref{alg:benchmark} explains how the benchmark becomes a set of verifiable contracts rather than a loose collection of prompts. Line 1 constructs the spatial substrate for each city, including the vector, raster, network, and topology layers needed by later tasks. Line 2 converts those layers into machine-readable schema records, which are essential for detecting invented fields and CRS errors. Lines 3--4 enumerate task families and instantiate concrete variants with distances, $k$-nearest-neighbor settings, and output columns. Line 5 attaches the constraint vocabulary used by the later verifiers, so each natural-language task has an explicit executable contract.

\subsection{Constraint-Aware Prompting}

The \llmonly{} baseline receives the query, local input paths, dataset schemas, and the expected output path/columns. The \ours{} setting additionally exposes task constraints, expected methods, and forbidden methods in the initial prompt. This makes the prompt a compact interface between natural-language intent and machine-checkable GIS contracts. The prompt also instructs the model to preserve source objects with zero matches, avoid invented fields, use stable GeoPandas APIs, handle missing CRS metadata, and write the exact output path.

\subsection{Verification Layers}

\begin{figure}[t]
  \centering
  \includegraphics[width=\linewidth]{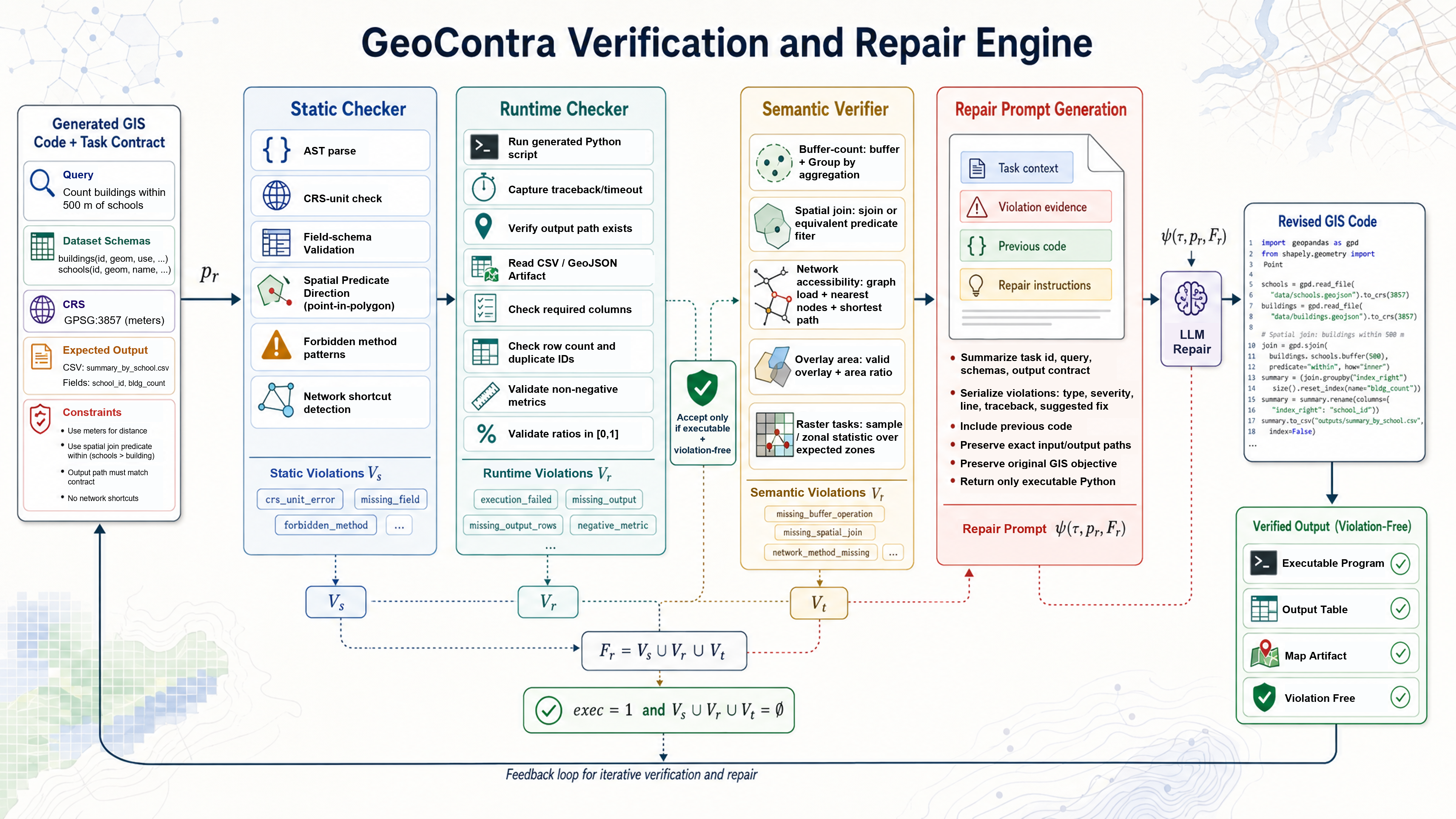}
  \caption{Detailed verification and repair engine. The static checker inspects AST structure, CRS-unit handling, field-schema consistency, spatial predicate direction, topology-sensitive operation patterns, forbidden method patterns, and network shortcuts. The runtime checker executes the generated program and validates tracebacks, timeouts, output-file existence, readable CSV/GeoJSON artifacts, required columns, row counts, duplicate identifiers, non-negative geographic metrics, and bounded ratios. The semantic verifier checks task-specific GIS operations such as buffer aggregation, spatial joins, graph-based shortest paths, overlay areas, and raster sampling or zonal statistics. Repair prompt generation then serializes task context, violation type, severity, line or traceback evidence, previous code, and hard path/objective constraints into the next repair prompt.}
  \label{fig:verify}
\end{figure}

Figure~\ref{fig:verify} expands the verifier-and-repair block from Figure~\ref{fig:overview}. The static checker first reasons over source code and task metadata before execution, but its target is geographic validity: whether the program is likely to preserve CRS semantics, spatial-predicate direction, topology-sensitive operations, and method constraints. The runtime checker then turns execution into concrete artifact evidence, checking whether the output is readable and geographically plausible. The semantic verifier checks whether an executable artifact actually used the GIS operation implied by the task; finally, repair prompt generation packages all violation evidence into a targeted prompt for the next model call. This design separates four different failure surfaces: pre-execution geographic-rule violations, execution and output-contract errors, GIS-method omissions, and prompt-level repair control.

The static verifier is intentionally conservative. It parses the generated script and flags common GIS failure modes: metric operations before recognized projected or geodesic handling, spatial predicate direction mismatches, point-polygon predicate inversions, missing or invented fields, direct geometry-distance shortcuts in network tasks, and task-specific forbidden methods. The runtime verifier then executes the program and checks the actual artifact for geographic plausibility and reproducibility: the expected spatial units must be preserved, identifiers must not be duplicated or silently dropped, distances and travel times must be non-negative, and area ratios must stay within physically meaningful bounds. The semantic verifier adds lightweight task-level obligations, such as requiring an explicit spatial join or equivalent predicate filter for spatial-join tasks.

\begin{algorithmblock}{Verifier-Guided Generation and Repair}
\label{alg:repair}
\textbf{Input} & Task contract $\tau$, model $G_{\theta}$, maximum repair rounds $R_{\max}$.\\
1 & Generate initial program $p_0 = G_{\theta}(\phi(\tau))$.\\
2 & For $r = 0,\ldots,R_{\max}$, run static checks $V_s(p_r,\tau)$.\\
3 & Execute $p_r$ in the local project environment and collect runtime checks $V_r(p_r,\tau)$.\\
4 & Apply semantic verifier $V_t(p_r,\tau)$ to the final artifact and source code.\\
5 & If execution succeeds and $V_s \cup V_r \cup V_t=\emptyset$, return $p_r$.\\
6 & If $r=R_{\max}$ or no repair feedback is allowed, stop.\\
7 & Build a repair prompt from $\tau$, $p_r$, and the concrete violations, then sample $p_{r+1}$.\\
\textbf{Output} & Final program, output artifact, violation history, token usage, and correctness label.\\
\end{algorithmblock}

Algorithm~\ref{alg:repair} is the outer control loop that connects generation, verification, and repair. Line 1 obtains the initial program from the model using the task contract. Lines 2--4 run the three verifier layers in order: static source-code checks, runtime artifact checks, and semantic GIS-task checks. Line 5 defines the success condition as both executable and violation-free, which is stricter than ordinary code execution. Line 6 enforces the bounded repair budget. Line 7 invokes the repair-prompt construction in Algorithm~\ref{alg:repair-prompt}, ensuring that each subsequent model call receives concrete, local failure evidence.

\begin{algorithmblock}{Static Checker}
\label{alg:static}
\textbf{Input} & Task contract $\tau$, generated program $p$.\\
1 & Parse $p$ into an abstract syntax tree; if parsing fails, emit a syntax violation.\\
2 & Build a field catalog from dataset schemas, expected output columns, and common derived fields.\\
3 & Detect metric operations such as buffer, distance, area, or length before projected CRS or geodesic handling.\\
4 & Inspect spatial relations for predicate mismatch, point--polygon direction errors, and topology-sensitive misuse.\\
5 & Check network semantics and forbidden shortcut patterns such as direct geometry distance in network tasks.\\
6 & Flag references to fields outside the task schema unless they are locally created derived fields.\\
7 & Return static violations with severity, source line, message, and suggested fix.\\
\textbf{Output} & Static violation set $V_s(p,\tau)$.\\
\end{algorithmblock}

Algorithm~\ref{alg:static} formalizes the pre-execution geographic-rule checks. Line 1 makes the checker robust to syntax failures before deeper analysis. Line 2 creates the schema boundary that separates legitimate columns from hallucinated fields. Line 3 targets the most common GIS numerical error: treating longitude/latitude degrees as meters. Line 4 checks whether spatial predicates match both the task semantics, geometry direction, and topology-sensitive relation being requested. Line 5 prevents semantically invalid shortcuts, especially in network accessibility tasks where Euclidean distance does not answer a walking-network question. Line 6 catches field-name hallucinations, while line 7 packages each violation into repair-ready feedback.

\begin{algorithmblock}{Runtime Checker}
\label{alg:runtime}
\textbf{Input} & Task contract $\tau$, generated program $p$, execution result $e$.\\
1 & If execution is skipped, return an empty runtime set because no artifact is available.\\
2 & If execution fails or times out, emit an execution violation with traceback, stdout/stderr excerpt, or timeout message.\\
3 & Locate the expected output path from $\mathcal{Y}$ and require that the file exists.\\
4 & Read the output as a table or geospatial file; if reading fails, emit an unreadable-output violation.\\
5 & Check required columns, expected row-count rules, duplicate IDs, and empty outputs.\\
6 & Validate numeric contracts such as non-negative counts, distances, areas, lengths, times, and ratios in $[0,1]$.\\
7 & Return runtime violations with messages and concrete suggested fixes.\\
\textbf{Output} & Runtime violation set $V_r(p,\tau)$.\\
\end{algorithmblock}

Algorithm~\ref{alg:runtime} is the artifact-level checker that turns execution into evidence. Line 1 handles non-executed settings without inventing runtime evidence. Line 2 captures ordinary failures and timeouts with the details needed for repair. Line 3 enforces the exact output path from the task contract. Line 4 prevents silent success when a file exists but is malformed. Line 5 checks tabular obligations such as required columns, one-row-per-source rules, duplicate identifiers, and non-empty output. Line 6 checks metric plausibility, including non-negative distances and bounded ratios. Line 7 converts these failures into structured feedback for Algorithm~\ref{alg:repair-prompt}.

\begin{algorithmblock}{Semantic Verifier}
\label{alg:semantic}
\textbf{Input} & Task contract $\tau$, generated program $p$, execution result $e$.\\
1 & If $e$ is skipped or failed, defer to runtime feedback and return no semantic violation.\\
2 & Dispatch according to the task type recorded in $\tau$.\\
3 & For buffer-count tasks, require a buffer operation and aggregation back to source objects.\\
4 & For spatial-join tasks, require a spatial join or equivalent spatial predicate filter.\\
5 & For network-accessibility tasks, require graph-based nearest-node and shortest-path logic.\\
6 & For overlay and raster tasks, require overlay, raster read, sampling, or zonal-statistics logic consistent with the task contract.\\
7 & Return task-level violations that capture missing GIS operations not exposed by execution alone.\\
\textbf{Output} & Semantic violation set $V_t(p,\tau)$.\\
\end{algorithmblock}

Algorithm~\ref{alg:semantic} checks whether an executable script actually performs the requested GIS operation. Line 1 avoids duplicating runtime failures already handled elsewhere. Line 2 routes the program to a task-specific semantic template. Line 3 catches buffer-count scripts that write a table without buffering or aggregating. Line 4 catches spatial-join tasks that merely manipulate tables or geometries without a spatial predicate. Line 5 distinguishes true network analysis from direct Euclidean distance approximations. Line 6 extends the same principle to overlay and raster analysis, where the output must come from the correct spatial support and gridded representation. Line 7 returns concise semantic violations, which are especially useful for executable-but-wrong programs.

\begin{algorithmblock}{Repair Prompt Generation}
\label{alg:repair-prompt}
\textbf{Input} & Task contract $\tau$, previous program $p_r$, feedback violations $F_r \subseteq V(p_r,\tau)$.\\
1 & Summarize the task id, task type, query, dataset schemas, and expected output contract.\\
2 & Serialize $F_r$ with violation type, severity, message, source line or traceback, and suggested fix.\\
3 & Insert the previous program $p_r$ after the violation summary.\\
4 & Add hard instructions to preserve input/output paths, fix all listed violations, and keep the original task objective.\\
5 & Require the model to return only executable Python code without Markdown fences.\\
\textbf{Output} & Repair prompt $\psi(\tau,p_r,F_r)$ for sampling $p_{r+1}$.\\
\end{algorithmblock}

Algorithm~\ref{alg:repair-prompt} describes how \method{} turns verifier outputs into model-usable feedback. Line 1 reconstructs the task context so the model does not repair in isolation. Line 2 preserves the exact evidence from static, runtime, or semantic failures. Line 3 includes the previous code so the model can make a targeted patch rather than regenerate blindly. Line 4 prevents drift away from the original GIS objective or file contract. Line 5 keeps the response directly executable, which simplifies the next round of Algorithm~\ref{alg:repair}.

\subsection{Paired Evaluation}

The evaluation is paired at the task level: for each task and model family, we compare \llmonly{} and \ours{} under the same task contract and local data. This pairing allows us to attribute improvements to the contract and repair machinery rather than changes in the task distribution.

\section{Evaluation}
\label{sec:evaluation}

\subsection{Benchmark and Models}

The full benchmark contains 7,079 tasks from 15 Greater Boston areas. Table~\ref{tab:benchmark} summarizes the task distribution. The difficulty labels are easy (267 tasks), medium (5,310), and hard (1,502). Closed-model experiments evaluate DeepSeek-V4 and Kimi-K2.5 on the full benchmark under both settings, producing 14,158 runs per family. The open-model suite evaluates 11 retained models on 300 formal tasks under both settings, for 600 runs per model and 6,600 runs in total.

\textbf{Data construction.}
The 15 study areas are defined by bounding boxes around Cambridge, Somerville, Brookline, Chelsea, Watertown, Medford, Arlington, Malden, Everett, Revere, Belmont, Winchester, northern Newton, Allston--Brighton, and Jamaica Plain. For each area, we use OSMnx to download OpenStreetMap-derived features by semantic tags, including parks and gardens, schools, hospitals and clinics, bus stops, libraries, supermarkets, fire stations, subway stations, water bodies and waterways, and residential buildings. OSMnx is also used to build local walking graphs with \texttt{network\_type="walk"}, stored as GraphML together with edge GeoJSON layers. We then derive additional GIS layers from these raw sources: analysis units are created by clipping a regular projected grid to the area boundary; centroids are computed from parks, residential buildings, and analysis units; water-buffer zones are generated from water polygons and lines; and 50-meter GeoTIFF rasters are produced by rasterizing park cover, water cover, residential cover, public-service point counts, and combined green-blue or service indicators. To evaluate topology-aware reasoning, we also create controlled invalid and overlapping geometry layers from otherwise real spatial features. Each city has a machine-readable catalog recording paths, CRS, geometry types, fields, row counts, bounds, and graph/raster metadata.

\textsc{GeoContra-Real} is intended to be convincing in a setting where ready-made industrial GIS evaluation data are not publicly available. Real operational GIS data often contain private addresses, proprietary asset layers, inconsistent schemas, and unreleased task definitions, so industry workflows cannot simply be redistributed as a benchmark. We therefore construct a reproducible public-facing proxy with real city geospatial layers, local walk graphs, rasterized urban indicators, topology perturbations, and contract-level expected outputs. The benchmark is large enough to stress model reliability statistically, diverse enough to cover the main GIS operation classes, and concrete enough that each generated program can be executed and checked. This combination of scale, realism, task diversity, and verifiability is the core reason the benchmark supports stronger conclusions than a small hand-written prompt suite.

\begin{table}[H]
  \centering
  \caption{\textsc{GeoContra-Real} benchmark composition by task family and difficulty.}
  \label{tab:benchmark}
  \small
  \begin{tabular}{@{}lrrrrr@{}}
    \toprule
    Task family & Easy & Medium & Hard & Total & Share \\
    \midrule
    Buffer count & 0 & 3,006 & 0 & 3,006 & 42.5 \\
    Nearest neighbor & 0 & 1,395 & 0 & 1,395 & 19.7 \\
    Network accessibility & 0 & 0 & 1,325 & 1,325 & 18.7 \\
    Raster sampling & 0 & 444 & 0 & 444 & 6.3 \\
    Raster zonal statistics & 0 & 300 & 0 & 300 & 4.2 \\
    Spatial join & 267 & 30 & 0 & 297 & 4.2 \\
    Topology quality & 0 & 45 & 75 & 120 & 1.7 \\
    Overlay area & 0 & 90 & 15 & 105 & 1.5 \\
    Multi-step workflows & 0 & 0 & 87 & 87 & 1.2 \\
    \midrule
    \textbf{Total} & \textbf{267} & \textbf{5,310} & \textbf{1,502} & \textbf{7,079} & \textbf{100.0} \\
    \bottomrule
  \end{tabular}
\end{table}

\subsection{Metrics}

We report executable rate, spatial correctness rate, average static violations, average runtime violations, average semantic-verifier violations, repair rounds, runtime, and token usage. Spatial correctness is stricter than executability: a program must execute and pass all verifier layers. Percentage-point gains are computed as $\Delta_m$ in Section~\ref{sec:formulation}.

\subsection{Closed-Model Results}

\method{} substantially improves both closed-model families. Table~\ref{tab:closed-overall} shows that DeepSeek-V4 improves from \rate{47.6} to \rate{77.5} spatial correctness, a gain of $+29.9\pp$, while Kimi-K2.5 improves from \rate{57.7} to \rate{81.5}, a gain of $+23.8\pp$. Executability also increases: DeepSeek-V4 rises from \rate{90.9} to \rate{98.2}, and Kimi-K2.5 rises from \rate{94.7} to \rate{97.6}.

\begin{table}[H]
  \centering
  \caption{Closed-model overall results on 7,079 paired tasks per family.}
  \label{tab:closed-overall}
  \small
  \begin{tabular}{@{}llrrrrr@{}}
    \toprule
    Family & Setting & Exec. & Spatial & Static & Runtime & Repairs \\
    \midrule
    DeepSeek-V4 & \llmonly{} & 90.9 & 47.6 & 0.515 & 0.102 & 0.000 \\
    DeepSeek-V4 & \ours{} & 98.2 & 77.5 & 0.221 & 0.027 & 0.528 \\
    Kimi-K2.5 & \llmonly{} & 94.7 & 57.7 & 0.392 & 0.092 & 0.000 \\
    Kimi-K2.5 & \ours{} & 97.6 & 81.5 & 0.165 & 0.028 & 0.489 \\
    \bottomrule
  \end{tabular}
\end{table}

\begin{figure}[H]
  \centering
  \includegraphics[width=0.72\linewidth]{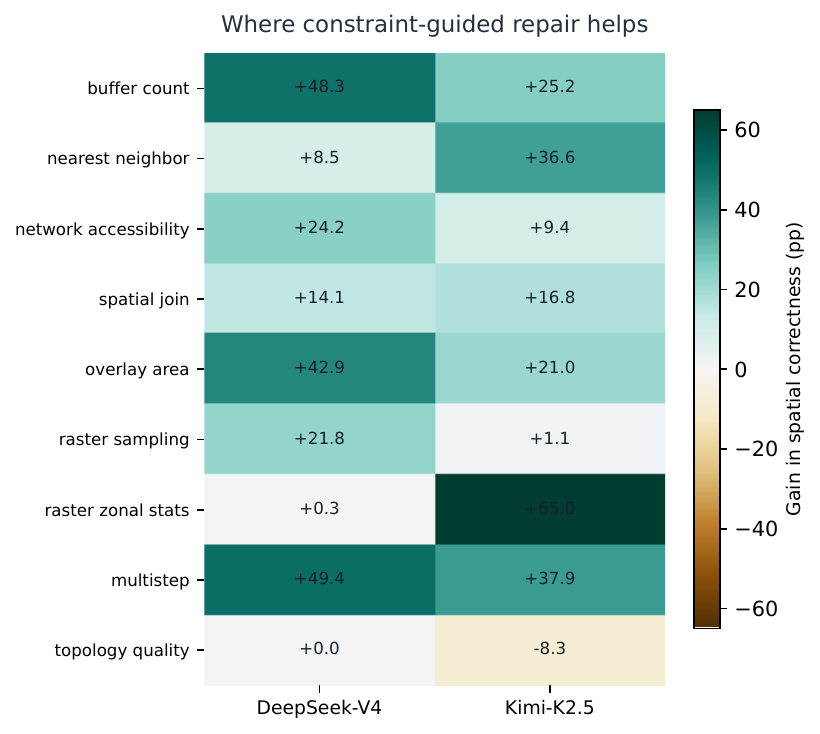}
  \caption{Spatial-correctness gains by task family and closed-model family. Each cell is \ours{} minus \llmonly{} in percentage points.}
  \label{fig:task-delta}
\end{figure}

The gains are not uniform, which is useful diagnostically. Figure~\ref{fig:task-delta} shows task-family gains. DeepSeek-V4 benefits most on multi-step workflows ($+49.4\pp$), buffer counts ($+48.3\pp$), and overlay area tasks ($+42.9\pp$). Kimi-K2.5 benefits most on raster zonal statistics ($+65.0\pp$), multi-step workflows ($+37.9\pp$), and nearest-neighbor tasks ($+36.6\pp$). The topology-quality family is the main non-improving pocket, suggesting that stronger topology-specific or reference-solution verifiers should be added in the next iteration.

\subsection{Open-Model Results}

\begin{figure}[H]
  \centering
  \includegraphics[width=0.92\linewidth]{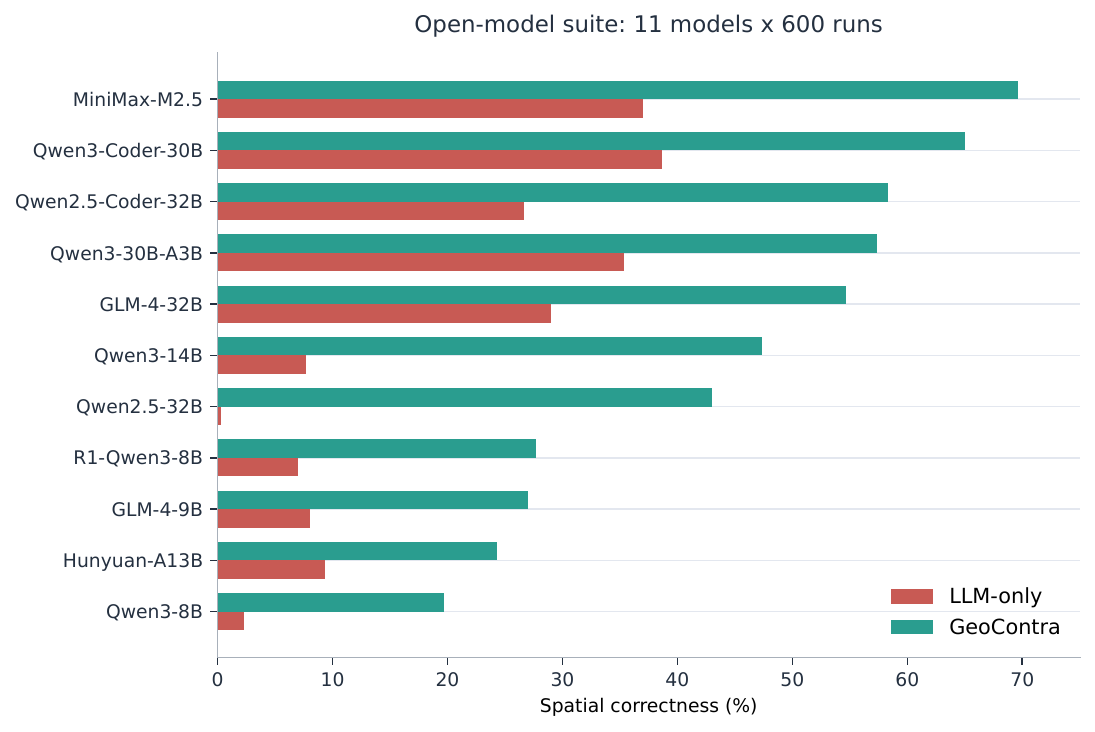}
  \caption{Open-model spatial correctness. Each model has 300 \llmonly{} and 300 \ours{} runs.}
  \label{fig:open-models}
\end{figure}

The 11-model open suite tests whether the mechanism generalizes beyond the two closed-model families. Average spatial correctness rises from \rate{18.3} under \llmonly{} to \rate{44.9} under \ours{}, a $+26.6\pp$ gain. Average executability rises from \rate{33.3} to \rate{61.7}. Table~\ref{tab:open-models} and Figure~\ref{fig:open-models} show that every retained open model improves.

The magnitude of improvement differs across model families because \method{} depends on two abilities: following a geography-constrained prompt and using verifier feedback to produce a syntactically executable repair. The largest gains occur when the baseline is weak but the model can still respond to structured constraints. For example, Qwen2.5-32B rises from \rate{0.3} to \rate{43.0} spatial correctness because its \llmonly{} setting almost never executes correctly (\rate{1.0} executability), while the contract-and-repair setting raises executability to \rate{61.3}. Qwen3-14B shows a similar pattern, rising from \rate{7.7} to \rate{47.3} spatial correctness as executability improves from \rate{22.0} to \rate{63.7}. By contrast, Hunyuan-A13B improves by only $+15.0\pp$ because many failures remain execution-level or API-level even after repair: its executability increases only from \rate{19.3} to \rate{30.0}, and runtime violations remain relatively high. Thus, \method{} is most effective when errors are contract-expressible and the model has enough coding ability to apply the repair; it is less effective when the model repeatedly produces non-executable GIS code or ignores low-level API constraints.

\begin{table}[H]
  \centering
  \caption{Open-model spatial correctness on 300 formal tasks per setting.}
  \label{tab:open-models}
  \small
  \begin{tabular}{@{}P{0.54\linewidth}rrr@{}}
    \toprule
    Model & \llmonly{} & \ours{} & Gain \\
    \midrule
    MiniMax-M2.5 & 37.0 & 69.7 & +32.7 \\
    Qwen3-Coder-30B-A3B & 38.7 & 65.0 & +26.3 \\
    Qwen2.5-Coder-32B & 26.7 & 58.3 & +31.7 \\
    Qwen3-30B-A3B & 35.3 & 57.3 & +22.0 \\
    GLM-4-32B & 29.0 & 54.7 & +25.7 \\
    Qwen3-14B & 7.7 & 47.3 & +39.7 \\
    Qwen2.5-32B & 0.3 & 43.0 & +42.7 \\
    DeepSeek-R1-Qwen3-8B & 7.0 & 27.7 & +20.7 \\
    GLM-4-9B & 8.0 & 27.0 & +19.0 \\
    Hunyuan-A13B & 9.3 & 24.3 & +15.0 \\
    Qwen3-8B & 2.3 & 19.7 & +17.3 \\
    \bottomrule
  \end{tabular}
\end{table}

\FloatBarrier

\subsection{Failure Diagnostics}

\begin{figure}[H]
  \centering
  \includegraphics[width=0.82\linewidth]{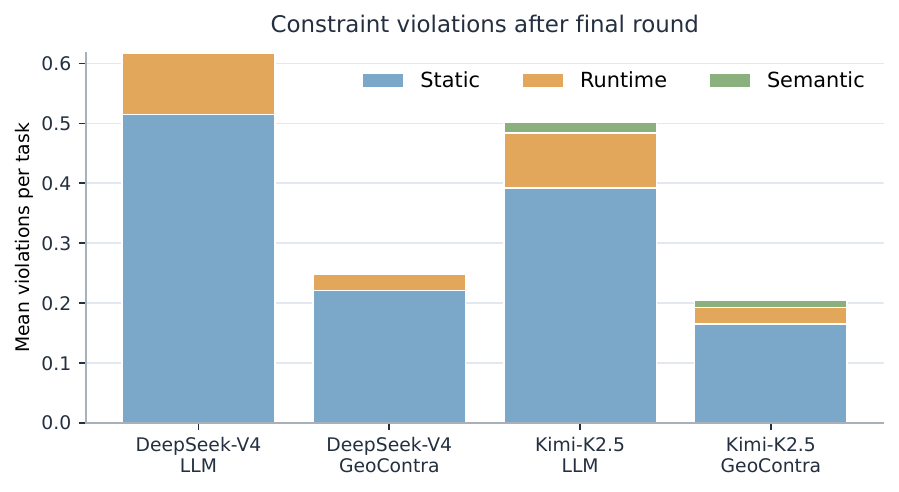}
  \caption{Average final violations per task after the last generation or repair round.}
  \label{fig:violations}
\end{figure}

The improvement is accompanied by a measurable reduction in final violations. Figure~\ref{fig:violations} decomposes average final violations into static, runtime, and semantic components for the two closed families. DeepSeek-V4's final runtime violations drop from 0.102 to 0.027 per task, and Kimi-K2.5's from 0.092 to 0.028. Static violations also decrease sharply: DeepSeek-V4 from 0.515 to 0.221 and Kimi-K2.5 from 0.392 to 0.165. This is important because many \llmonly{} outputs execute but still violate spatial contracts.

\FloatBarrier

\subsection{Case Studies}
\label{sec:case-studies}

We selected representative cases where \llmonly{} was incorrect and \ours{} was correct. Table~\ref{tab:cases} includes both non-executable failures and executable-but-wrong failures. The pattern is consistent: \method{} succeeds when the error can be expressed as a concrete violation, such as a forbidden shortcut, an invalid output cast, a missing spatial operation, or an impossible metric range.

\begin{table}[H]
  \centering
  \caption{Representative closed-model cases where \llmonly{} fails and \ours{} succeeds.}
  \label{tab:cases}
  \footnotesize
  \begin{tabular}{@{}P{1.6cm}P{3.0cm}P{0.27\linewidth}P{0.32\linewidth}@{}}
    \toprule
    Model & Task & \llmonly{} failure & \ours{} correction and interpretation \\
    \midrule
    DeepSeek-V4 &
    Watertown network accessibility to hospitals &
    The script executed, but the runtime checker found negative \texttt{nearest\_target\_min} values, violating non-negative travel-time semantics. &
    The constrained prompt produced a clean first-round program using the local GraphML, projected graph coordinates, \texttt{nearest\_nodes}, and shortest-path travel time. This shows the value of exposing network-method constraints before generation. \\
    DeepSeek-V4 &
    Revere multi-step residential access &
    The baseline mixed direct geometry distance into a network task and then failed execution. The static checker flagged the forbidden \texttt{.distance(} pattern. &
    Two repair rounds removed the forbidden shortcut, fixed a GeoPandas join issue, and produced a clean output. The explanation is local: verifier feedback names the exact illegal method and later the exact execution traceback. \\
    Kimi-K2.5 &
    Watertown bus-stop-to-analysis-unit assignment &
    The generated spatial join ran into a Pandas cast failure: string IDs such as \texttt{unit\_0011} were cast as nullable integers. &
    \method{} generated an executable point-in-polygon workflow with the expected string identifiers and output columns. The repair opportunity comes from treating output schema as a contract, not as a cosmetic formatting choice. \\
    Kimi-K2.5 &
    Somerville walk-edge intersection with water buffers &
    The baseline executed but the semantic verifier found no spatial join or equivalent predicate filter, so the result was not a valid intersection query. &
    One repair round produced a clean spatial-intersection workflow. This case illustrates why execution alone is insufficient: the wrong operation can still create a file. \\
    \bottomrule
  \end{tabular}
\end{table}
\FloatBarrier

\section{Discussion}
\label{sec:discussion}

The results suggest that LLM geospatial reliability depends less on model scale alone than on the structure surrounding the model. Smaller open models that fail frequently in the \llmonly{} setting can still benefit substantially from explicit contracts and verifier feedback. Conversely, strong closed models still make systematic GIS mistakes when the only target is plausible code. The largest improvements occur in tasks where errors are checkable without a full oracle: row preservation, CRS handling, spatial predicates, network-method use, and output schema validity.

From a GIScience perspective, the key implication is that trustworthy GeoAI should expose the geographic assumptions of an automated workflow as first-class computational objects. A contract specifies which spatial support is being analyzed, which spatial relation is intended, which CRS and unit system make measurements meaningful, which topology rules must hold, and which outputs can be interpreted as geographically plausible. This shifts LLM-based GIS from code assistance toward \emph{verifiable spatial analysis}: a model is not rewarded merely for generating an executable script, but for preserving the spatial question through the entire analysis chain.

The current verifier is deliberately lightweight. It does not yet compare every output against a reference solution, and some topology-quality tasks remain difficult. This is a design trade-off: the framework aims to scale to thousands of tasks while keeping checks interpretable. Future work should add stronger task-specific reference solvers, richer topology validators, uncertainty-aware repair policies that decide when additional rounds are worthwhile, and domain-specific geographic ontologies that distinguish acceptable analytical alternatives from violations of the task meaning.

A related limitation is the scope of the current semantic verifier. In this paper, the semantic layer mainly checks whether the generated workflow contains the GIS operation required by the contract, such as buffer construction, spatial joins, graph shortest paths, overlays, or raster sampling. This catches many executable-but-wrong programs, but it is still far from full spatial semantic understanding. It does not yet verify complex spatial-relation consistency across multiple layers, the spatial distributional plausibility of results, spatial autocorrelation patterns, neighborhood effects, or whether an output map ``looks'' geographically reasonable. These richer forms of semantic validation are important for future trustworthy GeoAI. They will likely require reference solvers for selected task classes, distribution-aware spatial diagnostics, topology-aware geometry comparison, and learned or rule-based checks over result maps and spatial patterns.

Some errors are also difficult for \method{} to repair. First, if a task is under-specified or if the contract does not encode the relevant geographic distinction, the verifier may be unable to tell whether two plausible interpretations are equivalent or not. Second, noisy or incomplete source data, such as missing OSM tags or ambiguous facility categories, cannot be fixed by program repair alone. Third, repeated low-level coding failures, unsupported library calls, package incompatibilities, or timeouts can exhaust the bounded repair budget before the model reaches a meaningful GIS workflow. Fourth, errors that preserve the required operation but produce an implausible spatial distribution may pass the current lightweight semantic checks. These cases clarify the boundary of the contribution: \method{} is a contract-grounded repair framework, not a complete geographic oracle.

\section{Conclusion}
\label{sec:conclusion}

We introduced \method, a geography-grounded verification and repair framework for LLM-driven GIS workflows. On 7,079 real tasks and 11 additional open models, the method consistently improves executability and spatial correctness while reducing final violations. The case studies show why: \method{} converts vague failure into actionable geospatial feedback, allowing the model to repair programs around CRS, schema, network, predicate, topology, and output-contract errors. More broadly, \method{} argues for a GIScience-centered path for GeoAI: automated GIS should be judged by whether it preserves geographic meaning, not only by whether it produces fluent code.

\section*{Data Availability Statement}

The benchmark data, task contracts, generated artifacts, evaluation outputs, and source code used in this study are publicly available in the GitHub repository \href{https://github.com/xxxyyyzzz3984/GeoContra-Real-Benchmark}{GeoContra-Real-Benchmark}. The repository provides the release snapshot and usage instructions to support reproduction, reuse, and follow-up research.

\section*{Statement on the Use of AI in Scientific Writing}
During the preparation of this work, the authors used ChatGPT (version 5.4) and Google Gemini to refine the manuscript's language and improve grammatical accuracy, as the authors are non-native English speakers. Additionally, these tools were employed to assist in the technical implementation and optimization of the research code. After using these services, the authors reviewed and edited the content as needed and take full responsibility for the originality, scientific integrity, and final version of the manuscript. The core research ideas, conceptual framework, and methodology were developed solely by the authors without the use of artificial intelligence.

\appendix
\section{Data Diversity of the Greater Boston Benchmark}
\label{app:data-diversity}

Although \textsc{GeoContra-Real} is built from Greater Boston, the 15 study areas cover substantially different urban morphologies. Dense inner-city areas such as Cambridge and Somerville have high walking-network and service-point densities; mixed residential and institutional areas such as Brookline, Allston--Brighton, and Jamaica Plain contain different balances of parks, schools, residential buildings, and water features; and lower-density suburban or coastal areas such as Winchester and Revere provide sparser road networks or stronger water-feature variation. The benchmark therefore stresses both compact urban structure and more suburban spatial patterns.

\begin{table}[H]
  \centering
  \caption{Diversity statistics across the 15 Greater Boston study areas. Densities are normalized by area in square kilometers.}
  \label{tab:appendix-diversity}
  \small
  \begin{tabular}{@{}lrrr@{}}
    \toprule
    Metric & Minimum & Median & Maximum \\
    \midrule
    Area (km$^2$) & 13.0 & 19.2 & 30.7 \\
    Tasks per area & 366 & 476 & 490 \\
    Walk-graph edges & 7,344 & 23,858 & 53,960 \\
    Walk-graph nodes & 2,652 & 8,585 & 19,616 \\
    Walk-edge density (edges/km$^2$) & 401.8 & 1,140.4 & 1,793.0 \\
    Bus-stop density (stops/km$^2$) & 3.0 & 11.2 & 18.9 \\
    Service-point density (points/km$^2$) & 4.2 & 17.0 & 26.0 \\
    Residential-building density (buildings/km$^2$) & 5.7 & 40.2 & 243.9 \\
    Park-feature density (features/km$^2$) & 1.0 & 3.2 & 10.1 \\
    Water polygons per area & 18 & 49 & 190 \\
    Water lines per area & 13 & 33 & 90 \\
    \bottomrule
  \end{tabular}
\end{table}

The range is important for generalization. For example, walk-edge density ranges from 401.8 edges/km$^2$ in Winchester to 1,793.0 in Cambridge, and service-point density ranges from 4.2 points/km$^2$ in Winchester to 26.0 in Somerville. Residential-building density varies even more sharply, from 5.7 buildings/km$^2$ in Revere to 243.9 in Cambridge. Water-related complexity is also heterogeneous, with water polygons ranging from 18 in Allston--Brighton to 190 in Revere. These differences make the benchmark more representative than a single homogeneous city core. At the same time, the appendix should not be read as claiming global coverage: extending the benchmark to grid-based western cities, rapidly urbanizing Asian cities, informal settlements, rural regions, and mountainous or coastal hazard contexts is an important next step.


\begin{thebibliography}{99}

\bibitem{janowicz2025geofm}
Janowicz, K., Gao, S., Mai, G., Hu, Y., and Zhu, R. (2025).
GeoFM: How will geo-foundation models reshape spatial data science and GeoAI?
\emph{International Journal of Geographical Information Science}.

\bibitem{natmi2025responsible}
Nature Machine Intelligence. (2025).
Towards responsible geospatial foundation models.
\emph{Nature Machine Intelligence}, 7, 395.

\bibitem{wu2025skysensepp}
Wu, Z., \emph{et al.} (2025).
SkySense++: A multi-modal remote sensing foundation model towards universal interpretation for Earth observation imagery.
\emph{Nature Machine Intelligence}, 7, 836--852.

\bibitem{wei2024geotool}
Wei, L., Li, G., and Gao, S. (2025).
GeoTool-GPT: A knowledge-based question-answering framework involving geospatial analysis tools.
\emph{International Journal of Geographical Information Science}, 39(3), 620--650.

\bibitem{ji2025georeasoning}
Ji, Y., Gao, S., Nie, Y., and Majic, I. (2025).
Revealing the impact of cross-domain knowledge on LLMs in understanding topological spatial relations in vector data.
\emph{International Journal of Geographical Information Science}.

\bibitem{hou2025geocogent}
Hou, S., \emph{et al.} (2025).
GeoCogent: An LLM-based agent for geospatial code generation.
\emph{International Journal of Geographical Information Science}.

\bibitem{lin2026geoagent}
Lin, X., \emph{et al.} (2026).
GeoAgent: A hierarchical LLM-based multi-agent architecture for autonomous spatial analysis.
\emph{International Journal of Geographical Information Science}.

\bibitem{deng2024k2}
Deng, C., \emph{et al.} (2024).
K2: A foundation language model for geoscience knowledge understanding and utilization.
\emph{Proceedings of the ACM Web Conference / WSDM Companion}.

\bibitem{zhang2024bbgeogpt}
Zhang, Y., \emph{et al.}(2024).
BB-GeoGPT: A framework for learning a large language model for geographic information science.
\emph{Information Processing \& Management}, 61(5), 103808.

\bibitem{mansourian2024chatgeoai}
Mansourian, A., \emph{et al.} (2024).
ChatGeoAI: Enabling geospatial analysis for the public through natural language, with large language models.
\emph{ISPRS International Journal of Geo-Information}, 13(12), 438.

\bibitem{zhang2024geogpt}
Zhang, Y., \emph{et al.} (2024).
GeoGPT: Understanding and processing geospatial tasks through an autonomous GPT.
\emph{International Journal of Applied Earth Observation and Geoinformation}, 131, 103976.

\bibitem{zhang2024mapgpt}
Zhang, Y., \emph{et al.} (2024).
MapGPT: An autonomous framework for mapping by integrating large language models and cartographic tools.
\emph{Cartography and Geographic Information Science}.

\bibitem{zhu2024flood}
Zhu, R., \emph{et al.} (2024).
A flood knowledge-constrained large language model interactable with GIS: When AI meets flooding.
\emph{International Journal of Geographical Information Science}, 38(11), 2180--2205.

\bibitem{gramacki2024evaluation}
Gramacki, P., \emph{et al.} (2024).
Evaluation of code LLMs on geospatial code generation.
\emph{Proceedings of the ACM SIGSPATIAL International Workshop on Geospatial Artificial Intelligence}.

\bibitem{zhang2024chatgpt4}
Zhang, W., and Gao, S. (2024).
Automating geospatial analysis workflows using ChatGPT-4.
\emph{Proceedings of the 2nd ACM SIGSPATIAL International Workshop on Advances in Urban-AI}.

\bibitem{hou2025geocodegpt}
Hou, S., \emph{et al.} (2025).
GeoCode-GPT: A large language model for geospatial code generation tasks.
\emph{International Journal of Applied Earth Observation and Geoinformation}, 141, 104456.

\bibitem{hou2025canllm}
Hou, S., \emph{et al.} (2025).
Can large language models generate geospatial code?
\emph{Geo-spatial Information Science}.

\bibitem{wu2025autogeevalpp}
Wu, C., \emph{et al.} (2025).
AutoGEEval++: A multi-level and multi-geospatial-modality automated benchmark for Google Earth Engine code generation.
\emph{Geo-spatial Information Science}.

\end{thebibliography}
\end{document}